\documentclass[journal]{IEEEtran}
\def\@IEEEclspkgerror{\ClassError{IEEEtran}}
\usepackage{enumitem}
\usepackage{cite}
\usepackage{graphicx}
\usepackage{algorithm}
\usepackage{algorithmic}
\usepackage{amsmath}
\usepackage{amssymb}
\usepackage{amsthm}

\usepackage{soul}
\usepackage[dvipsnames]{xcolor}
\usepackage{hyperref}
\usepackage[dvipsnames]{xcolor}

\makeatletter
\newcounter{parenttheorem}

\makeatother

\def\BibTeX{{\rm B\kern-.05em{\sc i\kern-.025em b}\kern-.08em
    T\kern-.1667em\lower.7ex\hbox{E}\kern-.125emX}}
\usepackage{lineno}
\begin{document}

\title{A Hybrid Model and Learning-Based Adaptive Navigation Filter}
\author{Barak Or, \IEEEmembership{Member, IEEE}, and Itzik Klein, \IEEEmembership{Senior Member, IEEE}
\thanks{Submitted: April 2022. First revision: June 2022. Second revision: July 2022. Accepted to IEEE TIM: August 2022. This is a preprint version.}
\thanks{Barak Or and Itzik Klein are with the Department of Marine Technologies, Charney School of Marine Science, University of Haifa, Haifa, 3498838, Israel (e-mail: barakorr@gmail.com, kitzik@univ@haifa.ac.il).}}

\markboth{A Hybrid Model and Learning-Based Adaptive Navigation Filter / Or and Klein - Preprint}%
{Or and Klein: A Hybrid Model and Learning-Based Adaptive Navigation Filter}
\maketitle

\begin{abstract}
The fusion between an inertial navigation system and global navigation satellite systems is regularly used in many platforms such as drones, land vehicles, and marine vessels. The fusion is commonly carried out in a model-based extended Kalman filter framework. One of the critical parameters of the filter is the process noise covariance. It is responsible for the real-time solution accuracy, as it considers both vehicle dynamics uncertainty and the inertial sensors quality. In most situations, the process noise covariance is assumed to be constant. Yet, due to vehicle dynamics and sensor measurement variations throughout the trajectory, the process noise covariance is subject to change. To cope with such situations, several adaptive model-based Kalman filters were suggested in the literature. 
In this paper, we propose a hybrid model and learning-based adaptive navigation filter. We rely on the model-based Kalman filter and design a deep neural network model to tune the momentary system noise covariance, based only on the inertial sensor readings. Once the process noise covariance is learned, it is plugged into the well-established model-based Kalman filter. 
After deriving the proposed hybrid framework, field experiment results using a quadrotor are presented and a comparison to model-based adaptive approaches is given. We show that the proposed method obtained an improvement of 25\% in the position error. Furthermore, the proposed hybrid learning method can be used in any navigation filter and also in any relevant estimation problem.
\end{abstract}


\section{Introduction}\label{sec:introduction}
\IEEEPARstart{A}{utonomous} vehicles, such as autonomous underwater vehicles (AUV) or quadrotors, are commonly equipped with an inertial navigation system (INS) and other sensors \cite{farrell2008aided} to provide real-time information about their position, velocity, and orientation \cite{wang2019constrained,wang2019enhanced,paull2013auv,gupte2012survey,xu2018enhancing,cui2021integrated}.
The INS has two types of inertial sensors; namely, gyroscopes and accelerometers. The former measures the angular velocity vector, and the latter measures the specific force vector. The inertial sensors are combined in an inertial measurement unit (IMU). Their use for navigation is challenging due to their high noise levels, which eventually accumulate, leading to a navigation solution drift over time. To solve this problem, an external aiding sensor, such as the global navigation satellite system (GNSS) receiver, is regularly used in a navigation filter \cite{wang2019constrained,wang2019enhanced,xu2018enhancing}. Such fusion is carried out using a nonlinear filter, where the error state implementation of the extended Kalman filter (es-EKF) is usually employed \cite{bar2004estimation,farrell2008aided}. In the filter, the process noise covariance matrix is based on the IMU measurements characteristics and/or vehicle dynamics while the external aiding measurements specification determines the measurement noise covariance matrix. Those two covariances matrices have a major influence on the filter performance. A common practice is to assume that both covariances are fixed during the vehicle operation. However, as the sensor characteristics are subject to change during operation and physical constraints do not allow capturing the vehicle dynamic optimally, those covariances should vary over time as the amount of uncertainty varies and is unknown. Thus, tuning the process or measurement noise covariances optimally can lead to a significant improvement in the filter performance \cite{bar2004estimation}. 

To cope with this problem, several model-based attempts have been made in the literature to develop optimal adaptive filters \cite{lyu2021adaptive,kang2016adaptive,davari2016asynchronous,jwo2007adaptive,jaradat2014enhanced,tong2017adaptive,he2021adaptive}. The most common among them is based on the filter innovation process \cite{mehra1970identification}, where a measure of the new information is calculated at every iteration, leading to an update of the process noise covariance matrix. Still, the question of the optimal approach to tune the system noise covariance matrix is considered open and is addressed in detail in \cite{zhang2020identification}. 

Recently, deep neural network (DNN) approaches were integrated in model based pedestrian dead-reckoning (PDR) algorithms. One of the initial works in the field is the robust IMU double integration, RIDI approach \cite{yan2018ridi}, where neural networks were trained to regress linear velocities from inertial sensors to constrain the accelerometer readings. Although it was the first work in the field, double integration is still needed and the error accumulates rapidly. In \cite{herath2020ronin}, the device orientation, in addition to the accelerometers and gyroscopes readings, was also used as input to a DNN architecture to regress the user velocity in 2D. The velocity is then integrated to obtain the user position. Later, PDRNet \cite{asraf2021pdrnet}, utilizes a smartphone location recognition classification network followed by a change of heading and distance regression network to estimate the used 2D position. One of the main challenges in model or learning based PDR is the estimation of the user walking direction as the device is not always aligned with the user motion. To cope with that, in \cite{manos2022walking} a novel DNN structure was designed for extracting the motion vector in the device coordinates, using accelerometer readings. More pedestrian inertial navigation examples are described by \cite{chen2020deep} where recent databases, methods and real-time inferences can be found. 

In AUV navigation, an end-to-end DNN approach was proposed to regress missing Doppler velocity log (DVL) beam measurements to provide the AUV velocity vector, only when a single beam is missing \cite{yona2021compensating}. Later, \cite{li2021underwater}, a DNN approach was used to address with a DVL failure scenarios to predict the DVL output. 

For land vehicle navigation a DNN model was presented to overcome the inaccurate dynamics and observation models of wheel odometry for localization, named RINS-W \cite{brossard2019learning}. Their novel approach exploited DNNs to identify patterns in wheeled vehicle motions using the IMU sensor only. To overcome the navigation solution drift in pure inertial quadrotor navigation, QuadNet, a hybrid framework to estimate the quadrotor’s three-dimensional position vector at any user-defined time rate was proposed  \cite{shurin2022quadnet}.  

Focusing on the estimation process, a self-learning square root-cubature Kalman filter was proposed \cite{shen2020seamless}. This hybrid navigation filter enhanced the navigation accuracy by providing observation continuously, even in GNSS denied environments, by learning transfer rules of internal signals in the filter. In \cite{or2022learning}, recurrent neural networks are employed to learn the vehicle’s geometrical and kinematic features to regress the process noise covariance in a linear Kalman filer framework.  In \cite{liu2021vehicle}, DNN based multi models (triggered by traffic conditions classification) together with an EKF was proposed to cope with GNSS outages.

Recently, a DNN model was proposed to solve the GNSS outages, mainly for unmanned air vehicles (UAV). In \cite{brossard2020denoising}, a convolutional neural network model was used for noise-free gyro measurements in open loop attitude estimation. They obtained state-of-the-art navigation performance in terms of attitude estimation where they compensated for gyro measurement errors as part of a strapdown integration approach. This trend of integrating learning approaches in various fields and applications also raises the motivation to adopt such approaches in the described problem.

In this paper, we propose a hybrid model and learning-based adaptive navigation filter. We rely on the model-based; yet, instead of using model-based adaptive process noise tuning,  we design a DNN to tune the momentary system noise covariance matrix, based only on the inertial sensor readings. Once, the process noise covariance is learned, it is plugged into the well-established, model-based es-EKF. To that end, we simulated many vehicle trajectories based on six baseline trajectories to create a rich dataset. This dataset contains rich dynamic scenarios with many motion patterns and different IMU noise covariances. This rich dataset was created to enable robustness to motion dynamics and IMU noise characteristics. Based on this dataset, we adopt a supervised learning (SL) approach to regress the process noise covariance. 

The main contributions of this paper are:
\begin{enumerate}
\item Derivation of an adaptive hybrid learning algorithm to momentary determine the process noise covariance matrix as a function of the IMU measurements. That is, instead of using model-based approaches (as been done in the last 50 years), to the best of our knowledge, we are the first to propose an adaptive hybrid learning algorithm.
\item Online integration of the proposed hybrid learning algorithm with es-EKF implementation of the navigation filter. In that manner, the proposed approach can be used with any external sensor aiding the INS.
\item The proposed approach principles can be further applied to any estimation problem that requires adaptive tuning of the process noise covariance.
\end{enumerate}

The advantages of the proposed approach lies in its hybrid fusion of the well celebrated es-EKF and applying learning approaches on the es-EKF soft spot, the adaptive determination of the process noise covariance. By using learning approaches we leverage from their ability to generalization of intrinsic properties appearing in sequential datasets and, therefore, better cope with varying conditions affecting the process noise values.

After deriving and validating the proposed framework using the stimulative dataset, field experiments using a quadrotor are done to evaluate the robustness and performance of the proposed approach testing phase and compare it to other adaptive model-based approaches. The recordings from the field experiment are used as a test dataset and also to examine our network capability for generalization. In our setup, we used GNSS position measurements to update the INS; however, the proposed approach is suitable for any aiding sensor and for any platform. 

The rest of the paper is organized as follows: Section II presents the problem formulation for the INS/GNSS fusion in an es-EKF implementation with constant and adaptive process noise covariance. Section III presents our proposed hybrid adaptive es-EKF. Section IV gives the results, and Section V presents the conclusions of this study.
\section{Adaptive Navigation Filter}
\label{sec:Adaptive}
The nonlinear nature of the INS equations requires a nonlinear filter. The most common filter for fusing INS with external aiding sensors is the es-EKF \cite{farrell2008aided}, and in particular, with a $15$ error states implementation:
\begin{equation}
\delta {\bf{x}} = {\left[ {\begin{array}{*{20}{c}}
{\delta {{\bf{p}}^n}}&{\delta {{\bf{v}}^n}}&{\delta {{\bf{\varepsilon }}^n}}&{{{\bf{b}}_a}}&{{{\bf{b}}_g}}
\end{array}} \right]^T} \in {{\mathbb{R}}^{15 \times 1}},
\end{equation}
where $\delta {{\bf{p}}^n} \in {{\mathbb{R}}^{3 \times 1}}$ is the position vector error states expressed in the navigation frame, $\delta {{\bf{v}}^n} \in {{\mathbb{R}}^{3 \times 1}}$ is the velocity vector error states expressed in the navigation frame, ${{\bf{\delta \varepsilon }}^n} \in {{\mathbb{R}}^{3 \times 1}}$ is the misalignment vector state expressed in the navigation frame, ${{\bf{b}}_a} \in {{\mathbb{R}}^{3 \times 1}}$ is the accelerometer bias residuals vector state expressed in the body frame, and ${{\bf{b}}_g} \in {{\mathbb{R}}^{3 \times 1}}$ is the gyro bias residuals vector state expressed in the body frame. The navigation frame is denoted by $n$ and the body frame is denoted by $b$. The navigation frame center is located at the body center of mass, where the $x$-axis points to the geodetic north, $z$-axis points down parallel to local vertical, and $y$-axis completes a right-handed orthogonal frame. The body frame center is located at the center of mass, $x$-axis is parallel to the longitudinal axis of symmetry of the vehicle pointing forwards, the $y$-axis points right, and the $z$-axis points down such that it forms a right-handed orthogonal frame. The linearized, error-state, continuous-time model is
\begin{equation} 
\delta {\bf{\dot x}} = {\bf{F}}\delta {\bf{x}} + {\bf{G}}\delta {\bf{w}},
\end{equation}
where $ {\bf{F}} \in {{\mathbb{R}}^{15 \times 15}}$ is the system matrix, ${\bf{G}} \in {{\mathbb{R}}^{15 \times 12}}$ is the shaping matrix, and $\delta {\bf{w}} = {\left[ {\begin{array}{*{20}{c}}
{{{\bf{w}}_a}}&{{{\bf{w}}_g}}&{{{\bf{w}}_{ab}}}&{{{\bf{w}}_{gb}}}
\end{array}} \right]^T} \in {{\mathbb{R}}^{12 \times 1}}$ is the system noise vector consisting of the accelerometer, gyro, and their biases random walk noises, respectively  \cite{bar2004estimation}. 
The system matrix, $\bf F$, is given by
\begin{equation}
{\bf{F}} = \left[ {\begin{array}{*{20}{c}}
{{{\bf{F}}_{pp}}}&{{{\bf{F}}_{pv}}}&{{{\bf{F}}_{p\varepsilon }}}&{{{\bf{0}}_{3 \times 3}}}&{{{\bf{0}}_{3 \times 3}}}\\
{{{\bf{F}}_{vp}}}&{{{\bf{F}}_{vv}}}&{{{\bf{F}}_{v\varepsilon }}}&{- {\bf{T}}_b^n}&{{{\bf{0}}_{3 \times 3}}}\\
{{{\bf{F}}_{\varepsilon p}}}&{{{\bf{F}}_{\varepsilon v}}}&{{{\bf{F}}_{\varepsilon \varepsilon }}}&{{{\bf{0}}_{3 \times 3}}}&{{\bf{T}}_b^n}\\
{{{\bf{0}}_{3 \times 3}}}&{{{\bf{0}}_{3 \times 3}}}&{{{\bf{0}}_{3 \times 3}}}&{{{\bf{0}}_{3 \times 3}}}&{{{\bf{0}}_{3 \times 3}}}\\
{{{\bf{0}}_{3 \times 3}}}&{{{\bf{0}}_{3 \times 3}}}&{{{\bf{0}}_{3 \times 3}}}&{{{\bf{0}}_{3 \times 3}}}&{{{\bf{0}}_{3 \times 3}}}
\end{array}} \right],
\end{equation}
where ${{\bf{T}}_b^n}$ is the transformation matrix between body and navigation frame and ${{\bf{F}}_{ij}} \in {{\mathbb{R}}^{3 \times 3}}$ can be found explicitly in the literature (see \cite{farrell2008aided,groves2008principles,brown1992introduction}). The shaping matrix is given by 
\begin{equation}
{\bf{G}} = \left[ {\begin{array}{*{20}{c}}
{{\bf{T}}_b^n}&{{{\bf{0}}_{3 \times 3}}}&{{{\bf{0}}_{3 \times 3}}}&{{{\bf{0}}_{3 \times 3}}}\\
{{{\bf{0}}_{3 \times 3}}}&{ - {\bf{T}}_b^n}&{{{\bf{0}}_{3 \times 3}}}&{{{\bf{0}}_{3 \times 3}}}\\
{{{\bf{0}}_{3 \times 3}}}&{{{\bf{0}}_{3 \times 3}}}&{{{\bf{0}}_{3 \times 3}}}&{{{\bf{0}}_{3 \times 3}}}\\
{{{\bf{0}}_{3 \times 3}}}&{{{\bf{0}}_{3 \times 3}}}&{{{\bf{I}}_{3 \times 3}}}&{{{\bf{0}}_{3 \times 3}}}\\
{{{\bf{0}}_{3 \times 3}}}&{{{\bf{0}}_{3 \times 3}}}&{{{\bf{0}}_{3 \times 3}}}&{{{\bf{I}}_{3 \times 3}}}
\end{array}} \right],
\end{equation}
where ${\bf I}_{3 \times 3}$ is an identity matrix. \\
The corresponding discrete version of the navigation model (for small step sizes), as given in (2), is 
\begin{equation}
\delta {{\bf{x}}_{k + 1}} = {{\bf{\Phi }}_k}\delta {{\bf{x}}_k} + {{\bf{G}}_{{k}}}\delta {{\bf{w}}_k}.
\end{equation} 
The transition matrix, ${{\bf\Phi }_k}$, is defined by a first order approximation as 
\begin{equation}
{{\bf{\Phi }}_k} \buildrel \Delta \over = {\bf{I}} + {{\bf{F}}}\Delta {t},
\end{equation}
where $k$ is a time index, $\Delta t$ is the time step size, ${\bf{F}}$ is the system matrix (obtained using $(3)$ with the estimated state vector at time $k$), and $\delta {\bf{w}}_k$ is a zero mean white Gaussian noise vector. The discretized process noise is given by
\begin{equation}
{\bf{Q}}^d_k = {\bf{G}}{{\bf{Q}}^c}{{\bf{G}}^T}\Delta {t},
\end{equation}
where $\bf G$ is defined in $(4)$ (using the current estimated state vector at time $k$), and ${\bf Q}^c$ is the continuous process noise matrix. \\
The discrete es-EKF, as described here, is used in the fusion process between the INS and the external measurements. Equations $(8)-(12)$ summarize the navigation filter equations\cite{farrell2008aided,groves2008principles}:
\begin{equation}
\delta {\bf{\hat x}}_k^ -  = {\bf 0}
\end{equation}
\begin{equation}
{\bf{P}}_k^ -  = {{\bf{\Phi }}_{k - 1}}{\bf{P}}_{k - 1} {{\bf{\Phi }}_{k - 1}}^T + {\bf{Q}}_{k - 1}^d
\end{equation}
\begin{equation}
{{\bf{K}}_k} = {\bf{P}}_k^ - {{\bf{H}}_k}^T{\left[ {{{\bf{H}}_k}{\bf{P}}_k^ - {{\bf{H}}_k}^T + {\bf{R}}_k^d} \right]^{ - 1}}
\end{equation}
\begin{equation}
\delta {\bf{\hat x}}_k  = {{\bf{K}}_k}\delta {{\bf{z}}_k}
\end{equation}
\begin{equation}
{\bf{P}}_k  = \left[ {{\bf{I}} - {{\bf{K}}_k}{{\bf{H}}_{k}}} \right]{\bf{P}}_k^ -,
\end{equation}
where $\delta {\bf {\hat x}}^-_k$ is the prior estimate of the error state and $\delta {\bf {\hat x}}_k$ is the posterior estimate, with their corresponding covariance matrices ${\bf P}_k^-$ $(9)$ and ${\bf P}_k$ $(12)$, respectively, ${\bf K}_k$ $(10)$ is the Kalman gain, $\delta {\bf z}_k$ is the $k$ measurement residual, ${\bf H}_k$ is the measurement matrix, and ${\bf R}_k$ is the measurement noise covariance matrix, assumed known and constant.
\subsection{Measurement Model}
\label{sec:Measurement}
The GNSS position measurements are available in a constant and lower frequency then that of the IMU.
After processing, the GNSS receiver outputs the vehicle position vector in the navigation frame, where it is used as an update in the navigation filter. Hence, the corresponding time-invariant GNSS measurement matrix is given by
\begin{equation}
{\bf{H}}_{GNSS} = \left[ {\begin{array}{*{20}{c}}
{{{\bf{I}}_{3 \times 3}}}&{{{\bf{0}}_{3 \times 12}}}
\end{array}} \right] \in {{\mathbb{R}}^{3 \times 15}}.
\end{equation}
The corresponding measurement residual is given by
\begin{equation}
\delta {\bf{z}}_{_{GNSS},j} = {{\bf{H}}_{GNSS}}\delta {{\bf{x}}_j} + {{\bf{\varsigma }}_{_{GNSS},j}},
\end{equation}
where ${{\bf{\varsigma }}_{_{GNSS},j}} \sim {\cal N}\left( {0,{{\bf{R}}^d}_{GNSS}} \right) \in {{\mathbb{R}}^{3 \times 1}}$ is an additive, discretized, zero mean white Gaussian noise. It is assumed that ${{\bf{\varsigma }}_j}$ and $\delta {\bf w}_j$ are uncorrelated.
\subsection{Model Based-Adaptive Noise Covariance}
\label{sec:ModelAdaptive}
The es-EKF minimizes the tracking error without changing $\bf Q$ matrix online; that is, the process noise covariance is constant throughout the operation. As shown in the literature, \cite{zhang2020identification} and the references therein, tuning $\bf Q$ online greatly improves the navigation filter performance.\\ 
For simplicity, we assume ${\bf Q}_k^d$ is a diagonal matrix, thus no correlation exists between the noise terms.

The optimal diagonal $\bf Q$ matrix at time step $k$ is defined by
\begin{equation}\label{eq:qstar}
{\bf{Q}}_k^* \buildrel \Delta \over = \arg \mathop {\min }\limits_{{{\bf{Q}}_k} \in {\cal Q}} \left\| {{{{\bf{\hat x}}}_k}\left( {{{\bf{Q}}_k}} \right) - {\bf{x}}_k^{GT}} \right\|_2^2,
\end{equation}
where $\cal Q$ is the admissible set of ${\bf Q}_k$ at time index $k$, ${\bf{x}}_k^{GT}$ is the ground truth (GT) state at time index $k$, and ${\left\|  \cdot  \right\|_2}$ is the second Euclidean norm.  \\
The most common approach to estimate $\bf Q$ in an adaptive es-EKF framework was suggested in \cite{mehra1970identification, zhang2020identification}, and is based on the innovation matrix for a window of size $\xi$:
\begin{equation}
{{\bf{C}}_k} \buildrel \Delta \over = \frac{1}{\xi }\sum\limits_{i = k - \xi  + 1}^k {{{\bf{\nu }}_i}{{\bf{\nu }}_i}^T},
\end{equation}
where ${{\bf{C}}_k}$ is the innovation matrix and the innovation vector is defined by
\begin{equation}
 {{\bf{\nu }}_k} \buildrel \Delta \over = {{\bf{z}}_k} - {\bf{H\hat x}}_k^ -
\end{equation}
The innovation matrix, (16), is used together with the Kalman gain, $\bf K$, to adapt $\bf Q$, as follows:
\begin{equation}
{\bf{\hat Q}}_k = {{\bf{K}}_k}{{\bf{C}}_k}{\bf{K}}_k^T.
\end{equation}
\section{A Hybrid Adaptive Navigation Filter}
\label{sec:Hybrid}
\begin{figure}[ht]
\centering
{\includegraphics[width=0.48\textwidth]{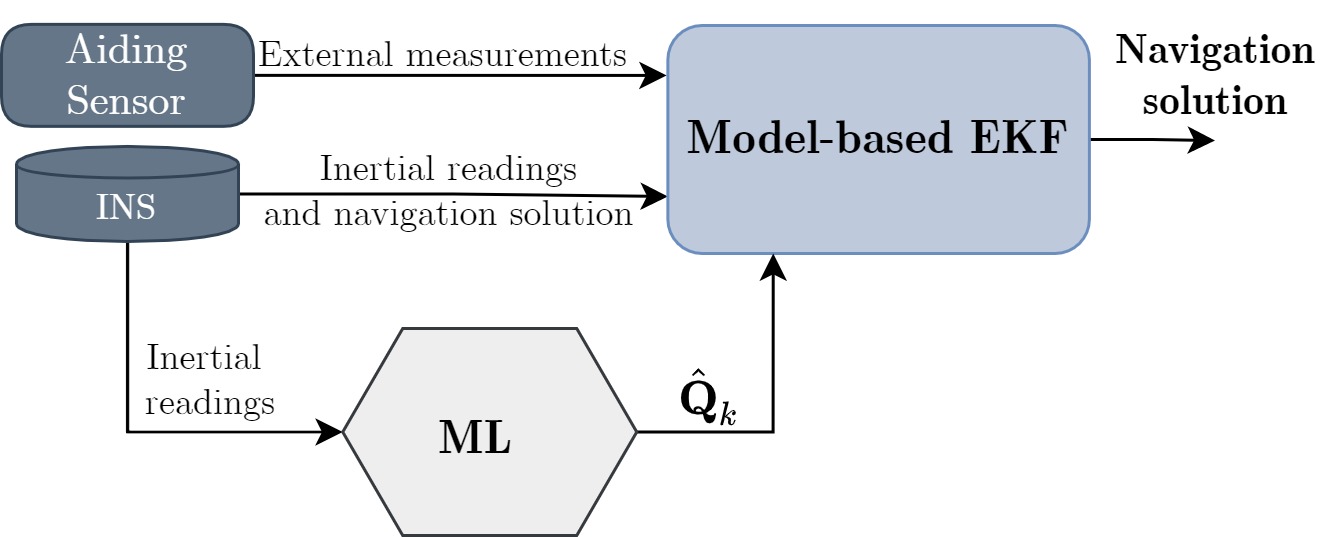}}
\caption{A hybrid adaptive navigation filter framework with online tuning of the process noise covariance matrix. }
\label{[Fig111}
\end{figure}

We rely on the well established, model-based es-EKF, yet instead of a model-based adaptive approach, a data-driven approach is used to regress the momentary process noise covariance, and, as a result, creates a hybrid adaptive es-EKF. To this end, the navigation filter equations (8)-(12) are used, but instead of using a constant process noise (7) or a model-based adaptive process noise (18), we derive a data-driven approach to estimate the time-varying continuous process noise covariance matrix ${\bf Q}^c$. Our proposed hybrid adaptive es-EKF framework, suitable for any aided INS scenario, is presented in Fig.~\ref{[Fig111}.
\subsection{Data-Driven Based Adaptive Noise Covariance}
\label{sec:DataDriven}
It is assumed that the continuous process noise covariance matrix, ${\bf Q}_k^c$, at time $k$, is a diagonal matrix with the following structure: 
\begin{equation}
\begin{array}{l}
{\bf{Q}}_k^c = \\
diag{\left\{ {q_f^{x*},q_f^{y*},q_f^{z*},q_\omega ^{x*},q_\omega ^{y*},q_\omega ^{z*},\varepsilon{{\bf{I}}_{1 \times 6}}} \right\}_k} \in{\mathbb{R}}{^{12 \times 12}},
\end{array}
\end{equation}
where the variance for each of the accelerometer axes is given by ${q_f^{i*}}$ and for the gyroscope by ${q_\omega ^{i*}}$ ($ i \in x,y,z$). The biases are modeled by random walk processes, and their variances are set to $\varepsilon=0.001$. By taking this approach, our goal is to ease a machine learning (ML) algorithm for accurately regressing the six unknown variances in (19) and thereby claim that these are the dominant variances needed in an adaptive framework.

We define a general one-dimensional series of length $N$ of the inertial sensor (accelerometer or gyroscope) readings in a single axis by 
\begin{equation}
{{\cal S}_k} = \left\{ {{{\bf{s}}_i}} \right\}_{i = k - N}^{k -1}.
\end{equation}
The discussed optimization problem is to find $q_k^*$ such that the position error is minimized. By doing that for all six IMU channels, the es-EKF considers an optimal system noise covariance matrix and provides improved tracking performance. The power of ML increases the ability to solve many difficult and non-conventional tasks. To determine ${\bf Q}_k^*$ in  \eqref{eq:qstar}, the problem is formulated as a SL problem. Formally, we search for a model to relate an instance space, ${\cal X}$, and a label space, ${\cal Y}$. We assume that there exists a target function, $\cal F$, such that ${\cal Y} = {\cal F}\left( {\cal X} \right)$. \\
Thus, the SL task for adaptive determination of the process noise covariance is to find ${\cal F}$, given a finite set of labeled examples, inertial sensor readings, and corresponding process noise variance values: 
\begin{equation}\label{eq:sets}
\left\{ {{{\cal S}_k},q_k^*} \right\}_{k = 1}^M.
\end{equation}
The SL approach aims to find a function ${\tilde {\cal F}}$ that best estimates ${\cal F}$. To that end, for the training process, a loss function, $l$, is defined to quantify the quality of ${\tilde {\cal F}}$ with respect to ${\cal F}$. The loss function is given by
\begin{equation}
{\cal L}\left( {{\cal Y},\hat {\cal Y}\left( {\cal X} \right)} \right) \buildrel \Delta \over = \frac{1}{M}\sum\limits_{m = 1}^M {l{{\left( {y,\hat y} \right)}_m}},
\end{equation}
where $M$ is the number of examples, and $m$ is the example index. Minimizing $\cal L$ in a training/test procedure leads to the target function. 
In our problem, the loss function is given by
\begin{equation}
{l_m} \buildrel \Delta \over = {\left( {q_m^* - {\hat q}_m} \right)}^2,
\end{equation}
where ${\hat q}_m$ is the estimated term obtained by the learning model during the training process.\\
In this work, the train dataset used to find the function ${\tilde {\cal F}}$ given the labeled examples \eqref{eq:sets}, as described in the next section.
\subsection{Dataset Generation}
\label{sec:dataset}
There are several existing inertial datasets as were recently summarized in \cite{shurin2022autonomous}. Yet, none of those datasets fits our problem as they do not provide enough IMU readings with different noise characteristics. As a consequence, we generated are own dataset using a simulation (detailed in Appendix \ref{sec:gen}).

\begin{figure*}[ht]
\centering
{\includegraphics[width=0.7\textwidth]{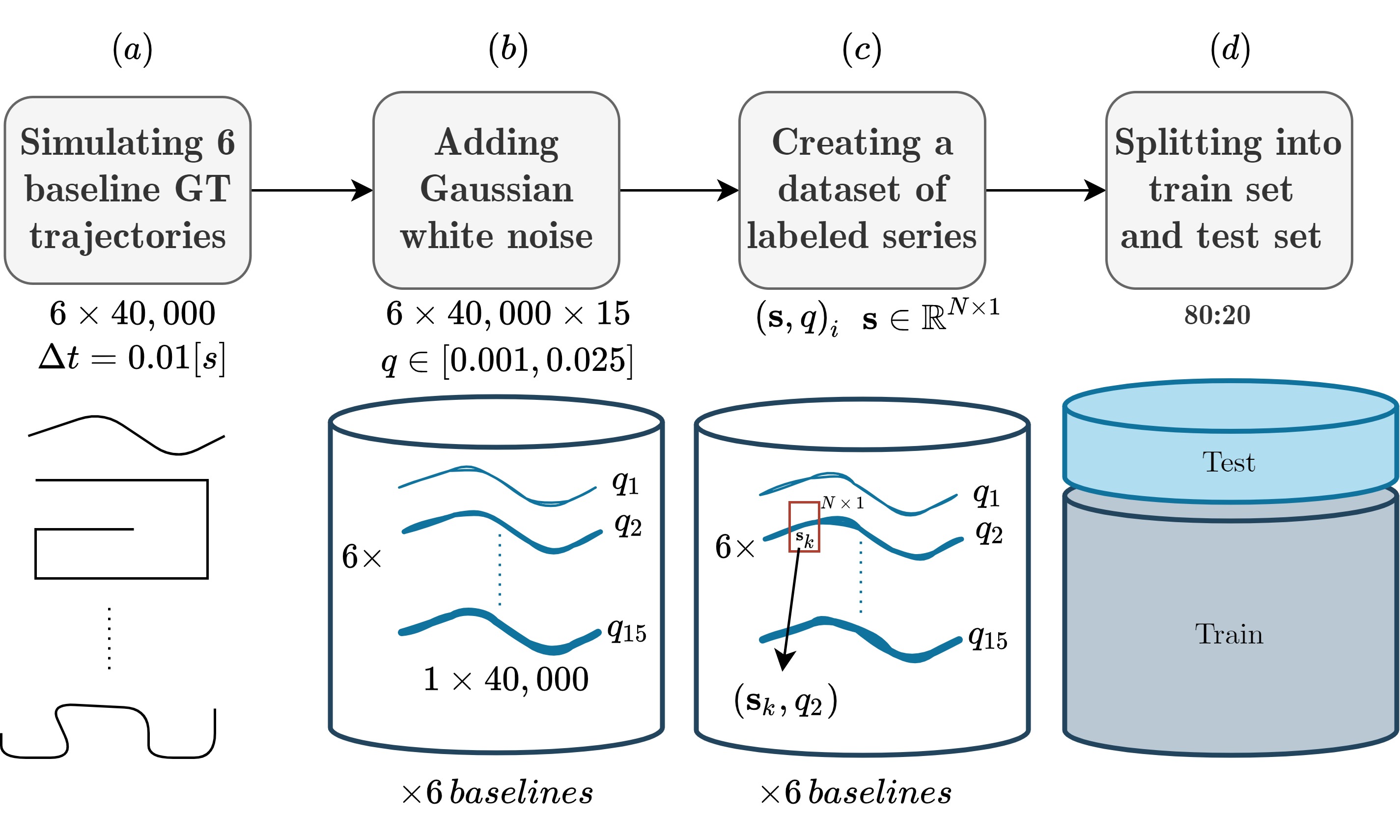}}
\caption{Dataset generation and pre-processing phase: (a) six different simulated baseline trajectories. Each baseline trajectory was created by tuning the IMU signals a period of $400[s]$ at a frequency of $100[Hz]$, resulting in a series of $6\times40,000$ samples for each baseline trajectory, obtained in phase (b). Example generation (c): given series length, $N$, examples are created and stored with their label. (d) split the database into train set and test set with a ratio of 80:20.}
\label{[Fig1}
\end{figure*}

\begin{figure*}[ht]
\centering
{\includegraphics[width=0.9\textwidth]{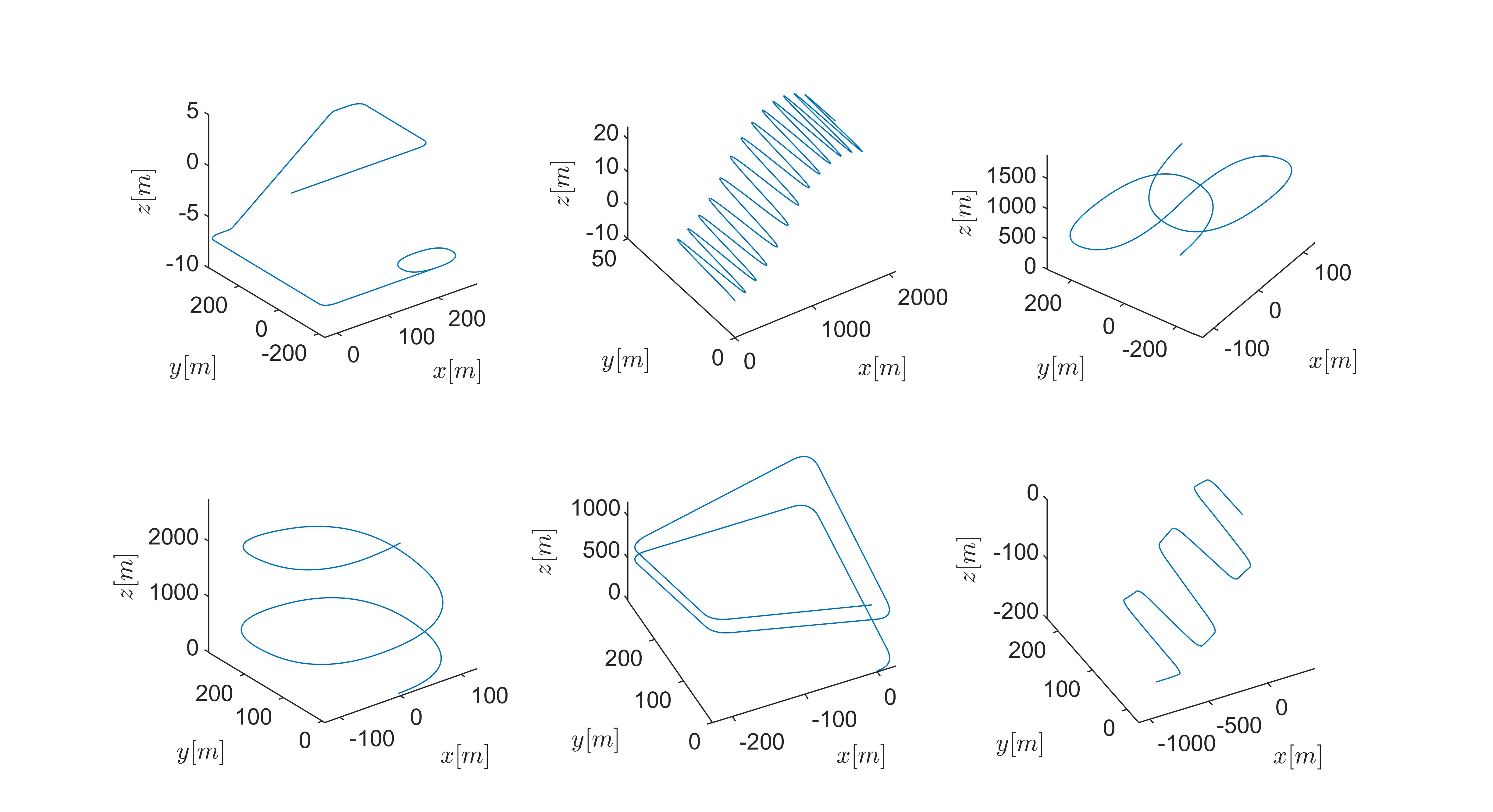}}
\caption{Our six baseline trajectories. }
\label{sixbaseline}
\end{figure*}

To generate the dataset six different baseline trajectories were simulated as presented in Fig. \ref{[Fig1}(a) and in Fig. \ref{sixbaseline}. The richness of the trajectories, as a result of their diversity, allows the establishment of a model to cope with unseen trajectories. Each baseline trajectory was created by generating ideal IMU readings for a period of $400$s in a sampling rate of $100$Hz, resulting in a sequence of $6\times40,000$ samples for each baseline trajectory as shown in Fig. \ref{[Fig1}.\\
Our task is to estimate the system noise covariance matrix, or specifically the noise variance of each IMU  channel. Hence, we divided each of the six IMU channels and corrupted  their perfect data with additive white Gaussian noise with variance in the range of $q\in[0.001,0.025]$, with 15 different values inside this interval, as shown in Fig. \ref{[Fig1}(b).\\
Hence, each baseline trajectory has $15$ series of $6\times40,000$ noisy inertial samples, as shown in Fig. \ref{[Fig1}(b). The justification to include a simple noise model lies in the momentary IMU measurement noise covariance sequence. A short time window for the IMU measurements is considered and allows characterizing the noise with its variance only. Next, the series length $N$ is chosen, and a corresponding labeled database is generated as shown in Fig. \ref{[Fig1}(c). Lastly, we choose $N=200$ samples and create batches corresponding to two seconds each with a total of $6\times200\times 15$. Then, these batches are randomly divided into train and test datasets in such a manner that all baseline trajectories are included in the train and the test sets in a ratio of 80:20, as described in Fig. \ref{[Fig1}(d). 
\subsection{Architecture design}
\label{sec:Architecture}
\begin{figure*}[h!]
\centering
{\includegraphics[width=1\textwidth]{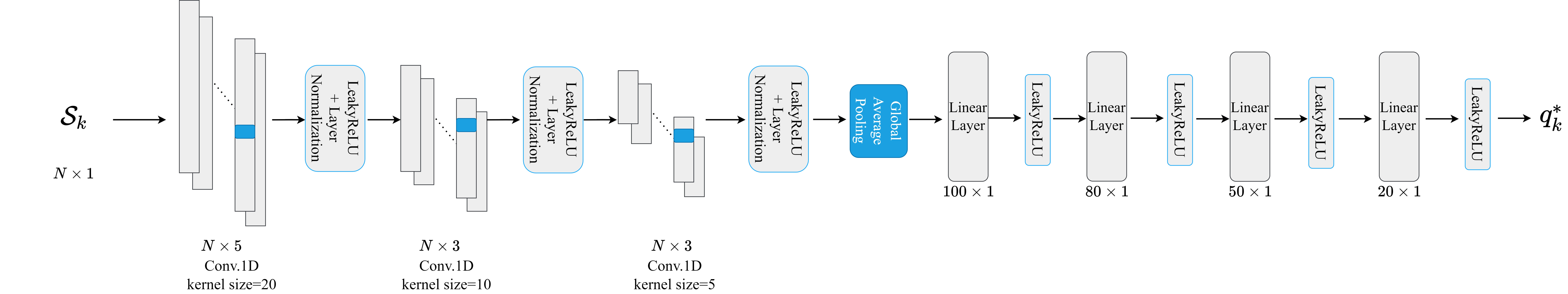}}
\caption{DNN architecture: Three Conv1D layers are followed by four linear layers and a global average pooling layer. Leaky ReLU activation functions are used after every layer to add nonlinearity and achieve better generalization capability.}
\label{[Fig3}
\end{figure*}

Several architectures were designed and evaluated. They included a recurrent DNN, and, in particular, the long short-term memory (LSTM) architectures for time series data. However, in our problem they did not perform well and failed in generalization of the noise variance property. One of the architecture we examined included a bi-directional LSTM layer followed by two linear layers. The training failed as the temporal information wasn't relevant for this task. On the other hand, an architecture with only convolution layers achieved satisfactory results, where it minimized the test loss very fast and obtained a low position root mean squared error (PRMSE) (the metric is defined in (25)). This result is due to the spatial information the noisy IMU signals have, allowing to capture the dynamics and statistical properties of the measurements. This information structure was easily captured by the spatial convolutional operates in the convolution layers. As the goal of this paper is to lay the foundations for a hybrid adaptive navigation filter approach, we describe here only the DNN architecture as presented in Fig. \ref{[Fig3}. 
Our DNN model has the following layers: 
\begin{enumerate}
\item {\bf Linear layer}: This is the simplified layer structure that applies a linear transformation to the incoming data from the previous layer. The input features are received in the form of a flattened 1D vector and are multiplied by the weighting matrix. 
\item {\bf Conv1D layer}: A convolutional, one-dimensional (Conv1D) layer creates a convolution kernel that is convolved with the layer input over a single spatial dimension to produce a vector of outputs. During the training phase the DNN learns the optimal weights of the kernels. In our DNN model the input layer is followed by a chain of three Conv1D layers. There are five kernels for the first layer and three kernels for the second and third layers. The kernel size decreases as the inputs go deep: the first Conv1D layer has twenty kernels, the second Conv1D layer has ten kernels, and the third Conv1D layer has five kernels.
\item {\bf Global average pooling layer}: Pooling layers help with better generalization capability as they perform down sampling feature map. In this way, the robustness of the DNN grows. For example, better handling with changes in the position of the features inside the 1D input vector. For the DNN architecture we selected the global average pooling method, which calculates the average for each input channel and flattens the data so it passes from Conv1D through the linear layers. 
\item {\bf Leaky ReLU}: Leaky rectified linear unit \cite{maas2013rectifier} is a nonlinear activation function obtaining a value $\alpha$ with the following  output:
\begin{equation}
f\left( \alpha  \right) = \left\{ {\begin{array}{*{20}{c}}
\alpha &{\alpha  > 0}\\
{0.01\alpha }&{\alpha  \le 0}
\end{array}}. \right.
\end{equation}
The main advantage of the leaky ReLU over the classical ReLU (0 for $\alpha \le 0$) is the small positive gradient when the unit is not active, which deals better with the dying ReLU problem \cite{lu2019dying}. The leaky ReLU functions were combined after each layer in the DNN model.
\item {\bf Layer normalization}: One of the challenges of DNN is that the gradients with respect to the weights in one layer 
are highly dependent on the outputs of the neurons in the previous layer, especially if these outputs change in a highly correlated way \cite{ba2016layer}. Batch normalization, and in particular layer normalization, was proposed to reduce such an undesirable covariate shift. The layer normalization was added after every Conv1D layer (together with the leaky ReLU layer). 
\end{enumerate}

The parameters of the above layers are as follow: a series of $200 \times 1$ samples is inserted to a one-dimensional convolutional layer (Conv1D) with a kernel size equals to $20$ with five filters. Then, a leaky ReLU activation function and normalization layer are applied, adding nonlinearity for better generalization capability of the DNN. We duplicated this structure twice more with three filters each and smaller kernels: the second with a kernel size equals to ten and the third with a kernel size equals to five. The output is inserted  to a global average pooling layer followed by four linear layers, with leaky ReLU activation functions between them, with $100,80,50$ and $20$ weights, retrospectively. Finally, the DNN outputs the process noise variance.

\subsection{Hybrid Learning Approach}
\label{sec:HybridScheme}
\begin{figure}[ht]
\centering
{\includegraphics[width=0.48\textwidth]{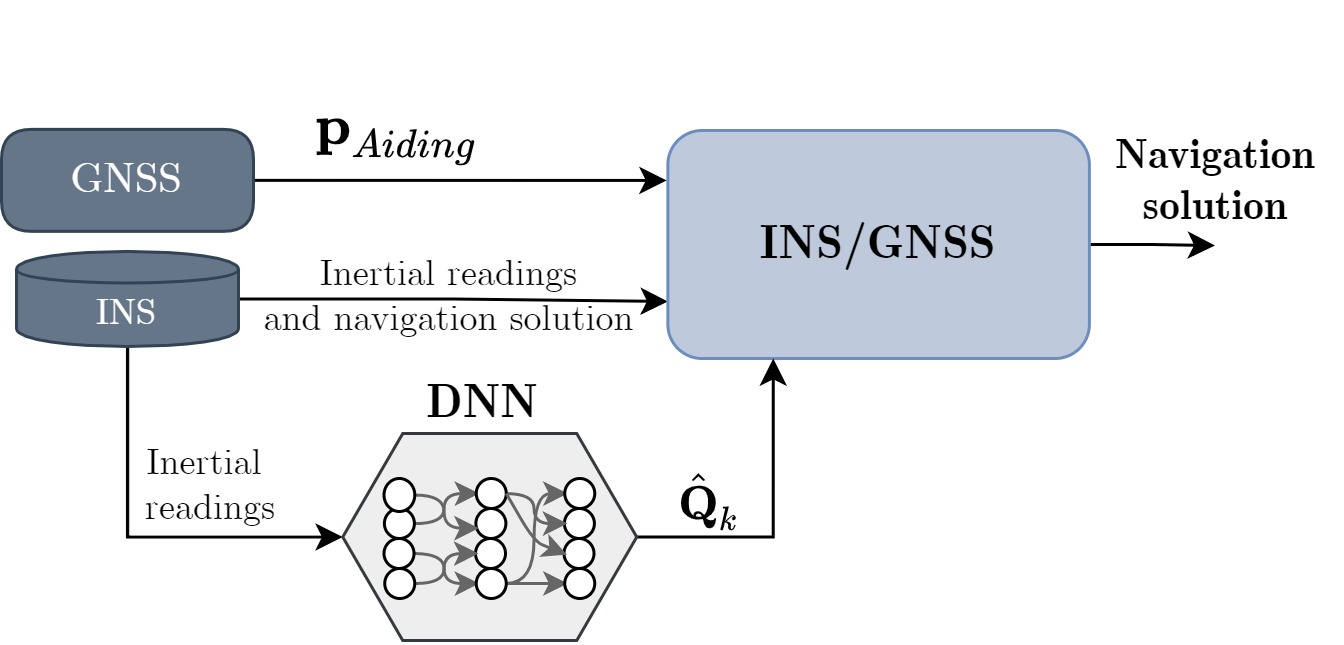}}
\caption{
Our adaptive hybrid adaptive filter applied to the INS/GNSS fusion process to tune the process noise covariance matrix, $\hat {\bf  Q}_k$. Both IMU and GNSS measurements are plugged into the INS/GNSS filter, while the six IMU channels are also inserted into the DNN for predicting the process noise covariance matrix.}
\label{[Fig5}
\end{figure}

In this work, GNSS position updates are employed to aid the INS and demonstrate our hybrid learning approach.
Applying the suggested adaptive tuning approach in online setting involves integrating the INS/GNSS with the regressor as presented in Fig. \ref{[Fig5}. Algorithm 1 describes the INS/GNSS with process noise covariance learning in a real time setup. The IMU signals are inserted both into the INS/GNSS filter $(8)-(12)$ and the DNN model. The regressed continuous  process noise covariance is plugged into  $(7)$ to obtain the discrete one, which in turn is substituted into $(9)$. 
\begin{algorithm}[h!]
 \caption{Hybrid adaptive filter applied to INS/GNSS}
 \begin{algorithmic}[1]
 \renewcommand{\algorithmicrequire}{\textbf{Input:}}
 \renewcommand{\algorithmicensure}{\textbf{Output:}}
 \REQUIRE ${\bf \omega}_{ib},{\bf f}_b,{\bf{v}}_{Aiding},\Delta t_0, \Delta \tau,T,tuningRate$
 \ENSURE  ${\bf{v}}^n,{\bf \varepsilon}^n$
 \\ \textit{Initialization} : ${\bf{v}}^n_0,{\bf{\varepsilon}}^n_0$
 \\ \textit{LOOP Process}
   \FOR {$t = 0$ to $T$}
   \STATE obtain ${\bf \omega}_{ib},{\bf f}_b$ 
  \STATE solve navigation equations (3)
  \IF {$(mod (t,\Delta \tau)$=0)}
  \STATE obtain ${\bf{p}}_{GNSS}$ 
 \STATE  update navigation state using the es-EKF (8)-(14)
  \ENDIF
  \STATE Calculate DNN model and predict ${\bf Q}_{k+1}^c$. 
  \IF {$mod(t,tuningRate)=0$}
\STATE${\bf Q}_{k + 1}^c = {\tilde {\cal F}_{NN}}\left( {{\cal S}_k} \right)$
\ENDIF
  \ENDFOR
 \end{algorithmic}
 \end{algorithm}
\section{Analysis and Results} \label{sec:Results}
We employ the following two metrics to evaluate the position accuracy of the proposed approach:
\begin{itemize}
    \item Position root mean squared error ($2^{nd}$ norm) for all three axes:
\begin{equation}
PRMSE = \sqrt {\frac{1}{c}\sum\limits_{k = 1}^{c}  {\sum\limits_{j \in \left\{ {x,y,z} \right\}}^{} {\delta {{\hat p}_{jk}}^2} } } .
\end{equation}
    \item Position mean absolute error for all six IMU channels:
\begin{equation}
PMAE = \frac{1}{c}\sum\limits_{k = 1}^{c} {\sum\limits_{j \in \left\{ {x,y,z} \right\}}^{} {\left| {\delta {{\hat p}_{jk}}} \right|}}.
\end{equation}
\end{itemize}
where $c$ is the number of samples, $k$ is a running index for the samples, $j$ is the running index for the Cartesian coordinate system, and $\delta {\hat p}_{jk}$ is the position error term.
\subsection{Train Dataset Analysis}\label{sec:DNNTraining}
\begin{figure}[ht]
\centering
{\includegraphics[width=0.48\textwidth]{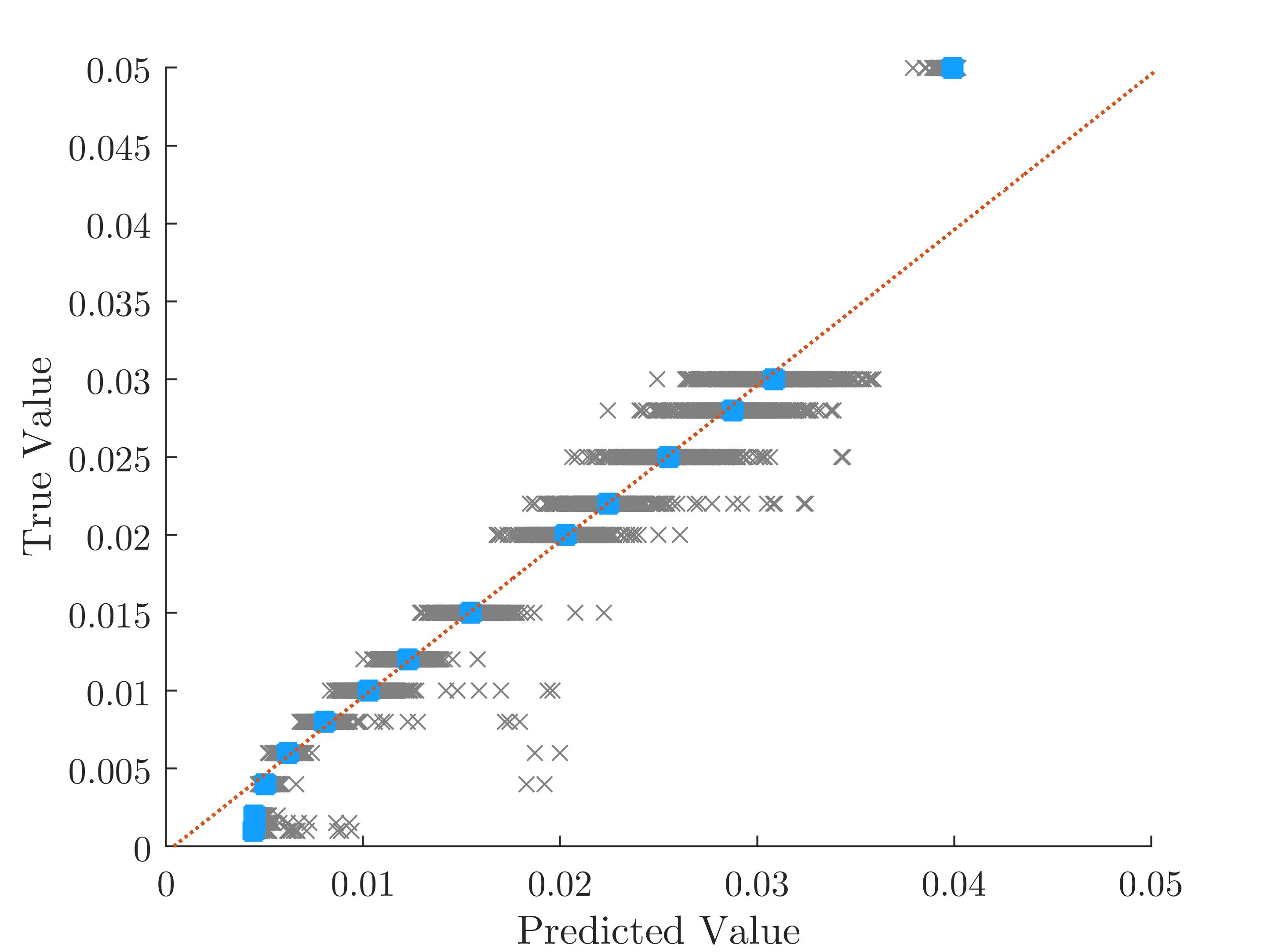}}
\caption{GT values versus predicted DNN values on the test set. All values are presented by gray crosses. The blue mean points represent the mean for each value, and the dashed red line (true value=predict value) represents the DNN generalization capability. }
\label{[Fig6}
\end{figure}

\begin{figure}[ht]
\centering
{\includegraphics[width=0.48\textwidth]{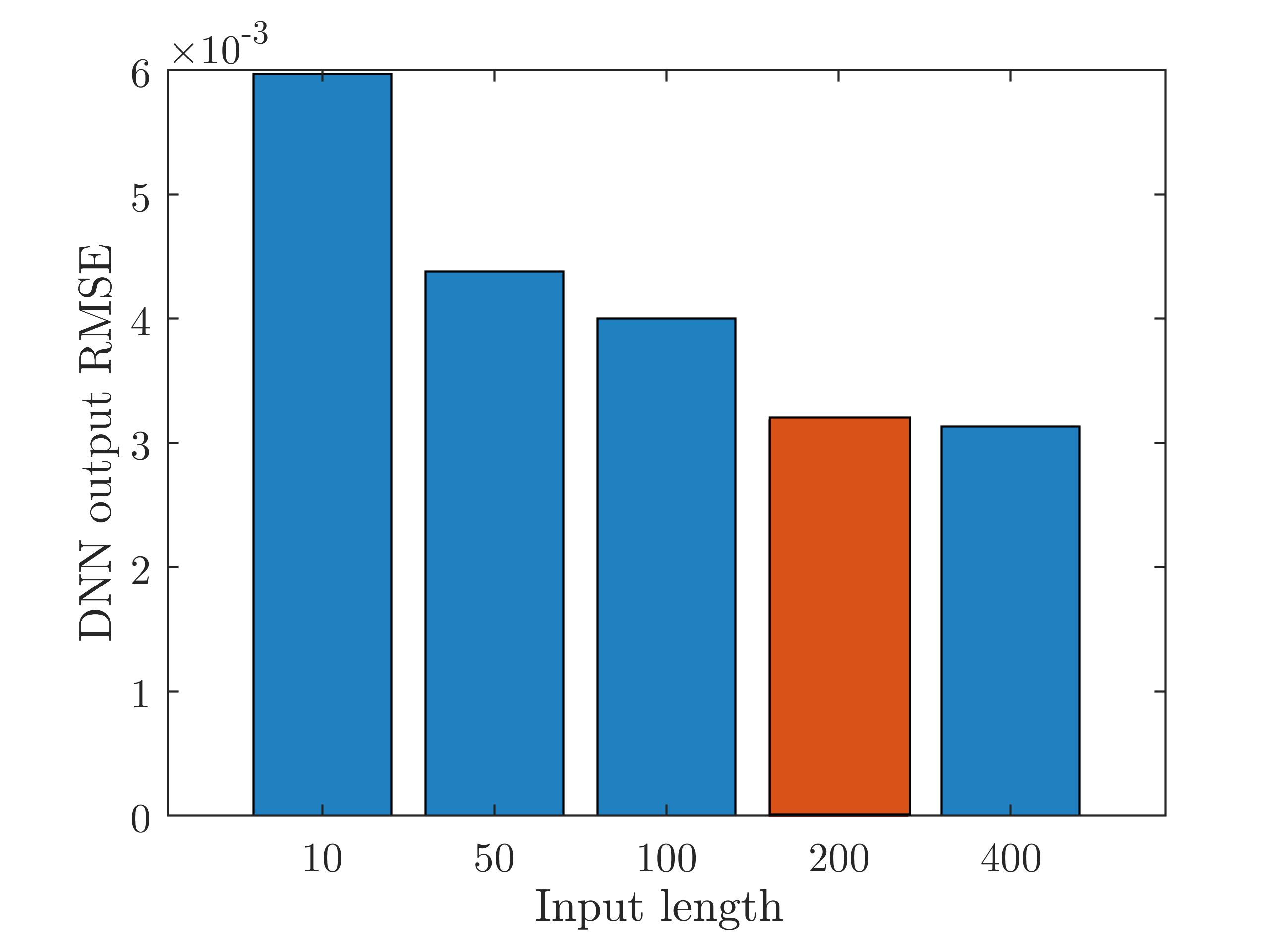}}
\caption{DNN RMSE vs. input length (N).}
\label{[Fig4}
\end{figure}

The proposed DNN architecture, presented in Section~\ref{sec:Architecture},  was trained on the dataset described in Section~\ref{sec:dataset}. After $30$ epochs it achieved an RMSE of $0.0032$ on the test set, showing the proposed DNN generalization capability and thereby establishing the trained DNN regressor model. The training process was done in mini-batches of $500$ examples each. The ADAM optimizer \cite{kingma2014adam} was used with an initial learning rate of $0.001$ and a learn-rate-schedule with a drop factor of $0.1$ after $20$ epochs. In our working environment (Intel i7-6700HQ CPU@2.6GHz 16GB RAM with MATLAB), the training time for input length of 200 samples elapsed about three hours. The averaged inference time is $0.02[s]$. Fig. \ref{[Fig6} presents the performance for $N=200$ samples where the red line is the desired one, representing when the GT value is equal to the DNN predicted value. The gray points represent the performance of the test set with the trained DNN model. The mean values of the test set lies near the red line (blue points) for most cases.\\
The INS/GNSS fusion is mostly performed under real-time conditions, where latency in the position computation might degrade the performance. Thus,  the  influence of the IMU sequence length input was examined. The system noise covariance matrix was learned, based on series of length $N$. As $N>>1$, the probability of learning the correct terms grows, since the DNN can capture the signal intrinsic properties easier using more data; however, the latency grows. Considering this trade-off, we trained the chosen architecture with various $N$ values and calculated the RMSE as shown in Fig.~\ref{[Fig4}. As expected, $N=400$ obtained the minimum RMSE, yet for the rest of the analysis we chose $N=200$ (similar RMSE) to receive the regression result as it is given in a shorter time period. For example, in the stimulative train and test dataset the sampling rate is $100$Hz, thus working with  $N=200$ gives the regression result every two seconds instead of working with four seconds ($N=400$). 
\subsection{Field experiment} \label{sec:experiment}
\begin{figure}[ht]
\centering
{\includegraphics[width=0.48\textwidth]{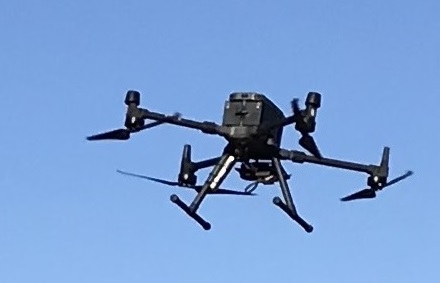}}
\caption{DJI matrice 300 in our field experiment.}
\label{[Fig7}
\end{figure}

\begin{figure}[ht]
\centering
{\includegraphics[width=0.48\textwidth]{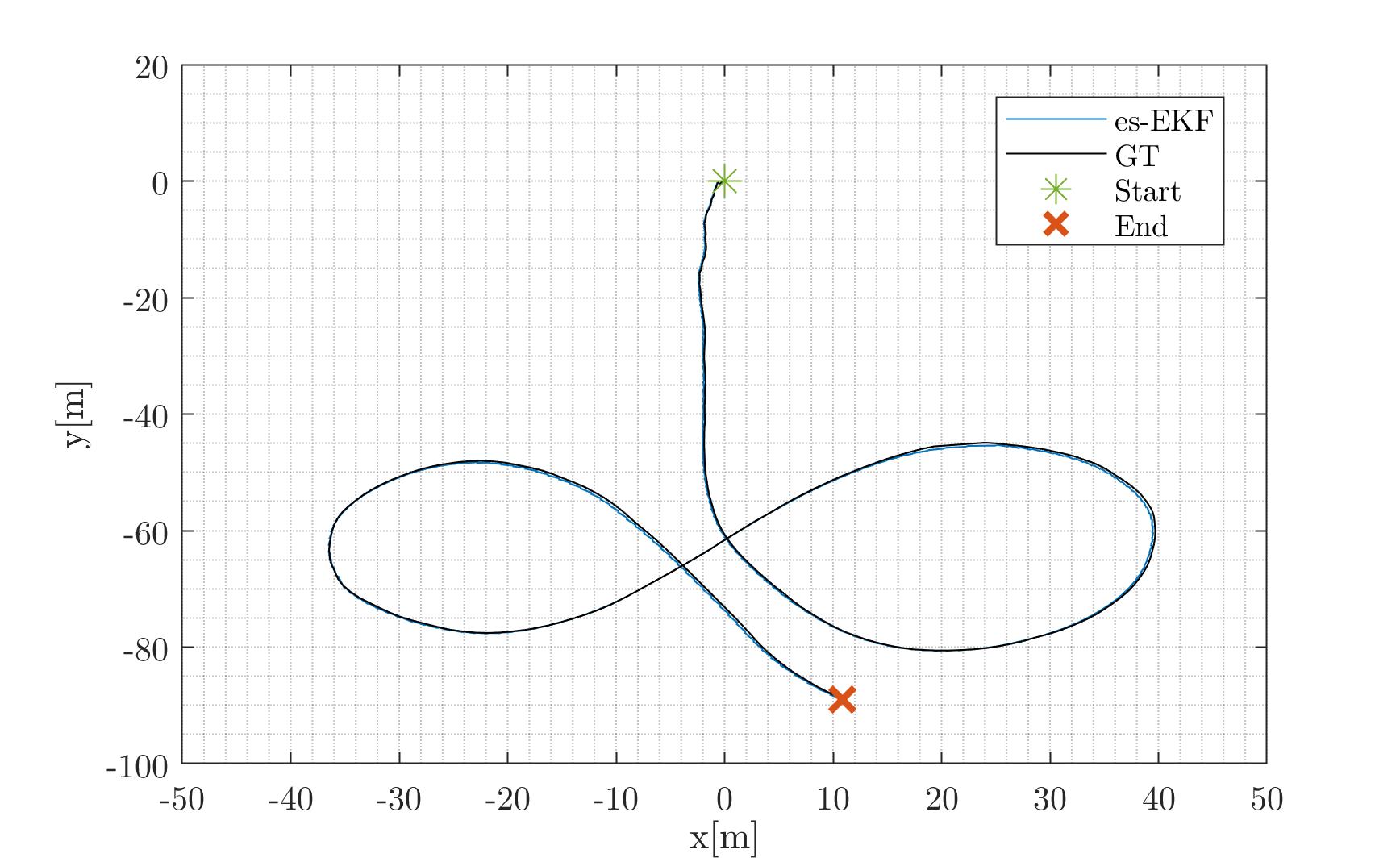}}
\caption{The trajectory of the quadrotor used to evaluate the proposed approach.}
\label{[Fig8}
\end{figure}

A field experiment with the DJI Matrice 300 \cite{DJI} was performed (Fig. \ref{[Fig7}). The trajectory is shown in Fig. \ref{[Fig8}. A figure-eight trajectory was made with some additional approximated straight line segments. The experiment parameters, including the GNSS RTK GT values and IMU, are given in Table I. The quadrotor dataset was published in \cite{dataset2022} and is publicly available at \url{https://github.com/ansfl/Navigation-Data-Project}.

\begin{table}[ht]
\caption {Experiment parameters} \label{tab:title}
\begin{center}
\begin{tabular}{ |c|c|c|c| } 
\hline
Description & Symbol & Value \\
\hline
GNSS noise (var) - horizon        & $R_{11},R_{22}$      &$0.01[m]^2$   \\ 
GNSS noise (var) - vertical        & $R_{33}$      &$0.02[m]^2$   \\ 
GNSS step size &$\Delta \tau$ & $0.2 [s]$   \\ 
IMU step size &$\Delta t$ & $0.001 [s]$   \\ 

Accelerometer noise (MPU-9250) & $q_{1:3}^*$   & $	0.02^2[m/s]^2$      \\ 
Gyroscope rate noise (MPU-9250) & $q_{4:6}^*$   & $
0.002^2[rad/s]^2$      \\ 
Experiment duration   & $T$   &$60 [s]$    \\
Initial velocity   & ${\bf{v}}^n_0$   &$[0, 0, 0]^T [m/s]$    \\
Initial position   & ${\bf{p}}^n_0$   &$[32.1^0, 34.8^0, 0]^T$    \\
\hline
\end{tabular}
\end{center}
\end{table}
Equations (2)-(12) were used as a model-based es-EKF with a constant process noise covariance matrix. Three different cases of the model-based es-EKF with constant process noise covariance matrix (MB-EKF-CQ) were examined:
\begin{itemize}
    \item \bf{MB-EKF-CQ1}: ${q_{1:3}} = {0.002},{q_{4:6}} = {0.02}$.
    \item \bf{MB-EKF-CQ2}: ${q_{1:3}} = {0.001},{q_{4:6}} = {0.01}$.
    \item \bf{MB-EKF-CQ3}: ${q_{1:3}} = {0.002},{q_{4:6}} = {0.01}$.
\end{itemize}
In addition, using (16)-(18), three different (window size length) cases of model-based  es-EKF with the adaptive process noise covariance matrix (MB-EKF-AQ) were examined: 
 \begin{itemize}
    \item \bf{MB-EKF-AQ1}: $\xi=1$.
    \item \bf{MB-EKF-AQ2}: $\xi=3$.
    \item \bf{MB-EKF-AQ3}: $\xi=5$.
\end{itemize}
All six approaches were compared to our proposed hybrid adaptive EKF (HB-AEKF) using the trained network, as described in the previous section. Results in terms of PMAE and PRMSE are presented in Table 2. The MB-EKF-AQ model-based adaptive filters obtained better performance than the MB-EKF-CQ constant process noise filters. In particular, MB-EKF-CQ3 achieved the best performance for constant process noise with a PMAE of $2.2[m]$ and a PRMSE of $1.6[m]$, where MB-EKF-AQ3 manged to improve them by $18\%$ and $19\%$, respectively. Our proposed hybrid adaptive approach, HB-AEKF, with an input length of $200$ samples, obtained the best performance, with $1.7[m]$ PMAE and $1.2[m]$ PRMSE improving MB-EKF-AQ3 by $5.5\%$ and $7.7\%$, respectively. 

\begin{figure}[h!]
\centering
{\includegraphics[width=0.48\textwidth]{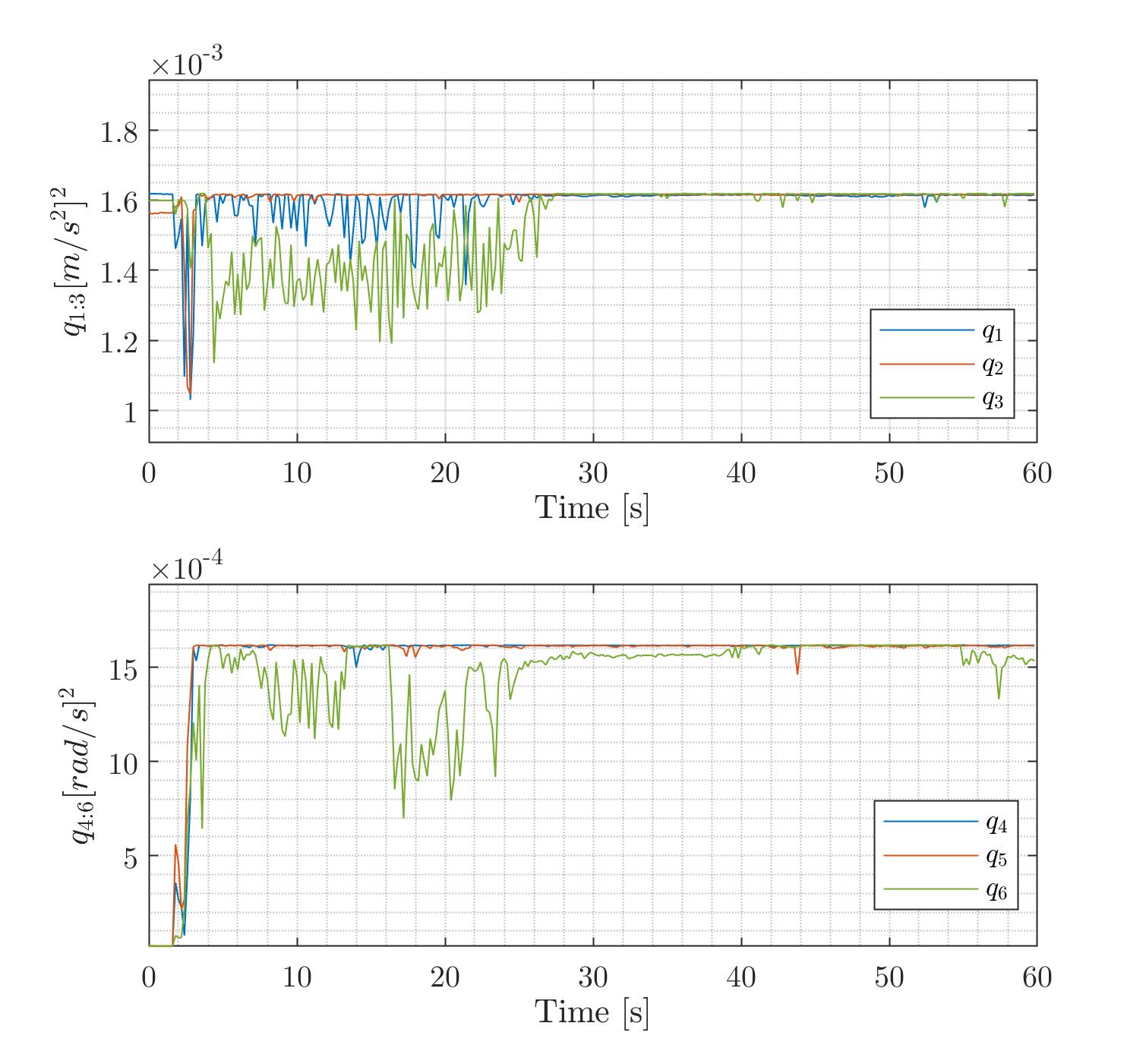}}
\caption{The process noise parameters as a function of time as obtained using our HB-AEKF approach. The upper plot shows the accelerometer noise variance for each axis as detected by the DNN model, and the lower plot shows the same for the gyroscope.}
\label{[Fig9}
\end{figure}

The $q$ parameters time evaluation were obtained using our HB-AEKF approach—are shown in Fig. \ref{[Fig9} for both accelerometers and gyroscopes. Their behavior shows fluctuations in the first $25$ seconds and then steady-state behavior. This behavior can be attributed to the convergence of the es-EKF $(8)$-$(12)$ once it obtains enough data and captures the IMU noise statistic. 

\begin{table}[h!]
\caption {Experiment results showing PMAE and PRMSE performance} \label{tab:title}
\begin{center}
\begin{tabular}{ |c|c|c| } 
\hline
$\bf Approach$ &$\bf  PMAE [m]$ &$\bf  PRMSE [m]$\\
\hline
HB-AEKF (ours)  & $\bf 1.7$  &$\bf 1.2$   \\ 
\hline
MB-EKF-AQ1  & $2.3$      &$1.6$   \\ 
\hline
MB-EKF-AQ2  & $2.0$      &$1.4$   \\ 
\hline
MB-EKF-AQ3  & $1.8$      &$1.3$   \\ 
\hline
MB-EKF-CQ1 & $2.4$      &$1.8$   \\
\hline
MB-EKF-CQ2  & $2.4$  &$1.7$   \\
\hline
MB-EKF-CQ3  & $2.2$      &$1.6$   \\
\hline
\end{tabular}
\end{center}
\end{table}

\begin{figure}[h!]
\centering
{\includegraphics[width=0.48\textwidth]{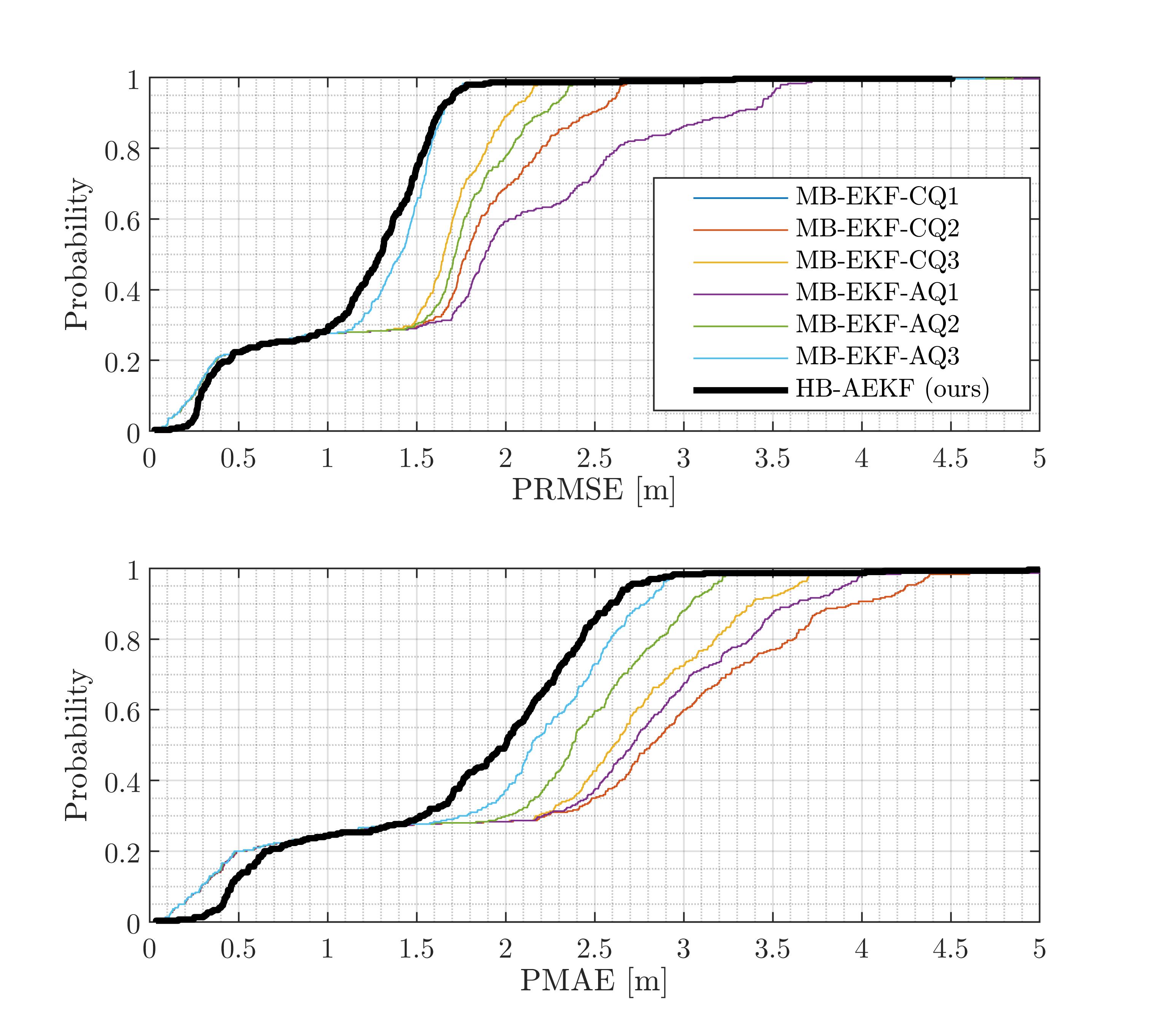}}
\caption{PRMSE and PMAE CDF graphs.}
\label{[Fig10}
\end{figure}

Fig. \ref{[Fig10} summarizes the positioning performance by providing two graphs of the cumulative density function (CDF); one representing PRMSE and the other the PMAE. In those plots, all three MB-EKF-AQ and three MB-EKF-CQ approaches are presented as well as our proposed HB-AEKF hybrid approach (black lines). Both graphs show clearly that the entire error is lower for HB-AEKF than all other approaches. The PRMSE of the MB-AEKF is lower than $1.7[m]$ and the PMAE lower than $2.8[m]$ for the entire scenario. In comparison with the MB-EKF-AQ1 approach, the maximum PRMSE and the maximum PMAE might grow too much: $3.7[m]$ and $4.0[m]$, respectively. 

\section{Conclusions}
\label{sec:Conc}
The proper choice for the system noise covariance matrix is critical for accurate INS/GNSS fusion. In this paper, a hybrid learning and model based approach was suggested in an adaptive filter framework.  To that end, a novel DNN-based approach to learn and  online tune the system noise covariance was developed. The input to the DNN is only the IMU readings. This learning approach was then combined with the model based es-EKF, resulting in a hybrid adaptive filter. 

The DNN model has been trained only once on a simulated database consisting of six different baseline trajectories. Analysis on  this dataset showed a trade-off between accuracy and the regression solution latency, where as the latency increases the accuracy decreases. Relying on this analysis, an input length of $200$ samples was chosen.

To validate the performance of the proposed approach, it was compared to six model-based approaches representing both constant and adaptive process noise covariance selection.  Quadrotor field experiments were conducted and used as an additional test dataset on the trained network. We obtained state-of-the-art results for each noise value compared to classical adaptive KF approaches; a PRMSE lower than $1.2[m]$ and a PMAE lower than $1.7[m]$ were obtained using our suggested approach, compared to a PRMSE of $1.6[m]$ and a PMAE of a $2.3[m]$ that were obtained using the innovation-based adaptive approach and classical tuning. Results in a total PRMSE improvement of $24\%$ and PMAE improvement of $27\%$. 

This improvement also demonstrates the robustness and generalization properties of the proposed DNN architecture, enabling it to cope with unseen data from different IMU sensors. In addition, and together with the train dataset results, it proves our hypothesis of regressing only the first six elements in the diagonal of the continuous process noise covariance matrix. 

Although demonstrated for quadrotor INS/GNSS fusion, the proposed approach can be elaborated for any external sensor aiding the INS and for any type of platform.

Although the present study demonstrated a small improvement over model-based methods using the test dataset  of a quadrotor INS/GNSS fusion, it introduces a hybrid framework to combine deep learning in an adaptive navigation filter. This approach can be applied to other platforms, dynamics and external sensor aiding the INS and serve as a foundation approach for future work in the field.

\appendix

\subsection{Abbreviations}
\begin{table}[h!]
\caption {Abbreviation and Description}
\begin{center}
\begin{tabular}{ |c|c| } 
\hline
Abbreviation & Description \\
\hline
AUV  & autonomous underwater vehicle \\
\hline
CDF  & cumulative density function \\
\hline
Conv1D  & one-dimensional convolution  \\
\hline
DNN  & deep neural network \\
\hline
DVL  &  Doppler velocity log \\
\hline
EKF & extended Kalman filter \\
\hline
es & error state \\
\hline
GNSS  & global navigation satellite system \\
\hline
GT  & ground truth \\
\hline
HB-AEKF  & hybrid based adaptive EKF \\
\hline
INS  & inertial navigation system \\
\hline
LSTM  & long short-term memory \\
\hline
MB-EKF-AQ  & model based EKF adaptive Q  \\
\hline
MB-EKF-CQ  & model based EKF constant Q  \\ 
\hline
ML & machine learning \\
\hline
NED  & North-East-Down \\
\hline
PDR  & pedestrian dead-reckoning \\
\hline
PMAE  & position mean absolute error \\
\hline
PRMSE  & position RMSE\\
\hline
RMSE  & root mean squared error\\
\hline
SL  & supervised learning \\
\hline
UAV  & unmanned air vehicles \\
\hline
\end{tabular}
\end{center}
\end{table}

\subsection{Dataset Generation}
\label{sec:gen}
\subsubsection{INS Equations of Motion}
The INS equations of motion include the rate of change of the position, velocity, and the transformation between the navigation and body frame, as shown in Fig.12. 

The position vector is given by
\begin{equation}
{{\bf{p}}^n} = {\left[ {\begin{array}{*{20}{c}}
\phi &\lambda &h
\end{array}} \right]^T} \in {\mathbb{R}^{3 \times 1}},
\end{equation}
where $\phi$ is the latitude, $\lambda$ is the longitude, and $h$ is the altitude. The velocity vector is Earth referenced and expressed in the North-East-Down (NED) coordinate system:
\begin{equation}
{{\bf{v}}^n} = {\left[ {\begin{array}{*{20}{c}}
{{v_N}}&{{v_E}}&{{v_D}}
\end{array}} \right]^T} \in \mathbb{R}^{3 \times 1},
\end{equation}
where $v_N,v_E,v_D$ denote the velocity vector components in north, east, and down directions, respectively. The rate of change of the position is given by \cite{farrell2008aided}
\begin{equation}
{{{\bf{\dot p}}}^n} = \left[ {\begin{array}{*{20}{c}}
{\dot \phi }\\
{\dot \lambda }\\
{\dot h}
\end{array}} \right] = \left[ {\begin{array}{*{20}{c}}
{\frac{{{v_N}}}{{{R_M} + h}}}\\
{\frac{{{v_E}}}{{\cos \left( \phi  \right)\left( {{R_N} + h} \right)}}}\\
{ - {v_D}}
\end{array}} \right],
\end{equation}
where $R_M$ and $R_N$ are the meridian radius and the normal radius of curvature, respectively.
The rate of change of the velocity vector is given by \cite{farrell2008aided}
\begin{equation}
{{{\bf{\dot v}}}^n} = {\bf{T}}_b^n{{\bf{f}}^b} + {{\bf{g}}^n} - \left( {\left[ {\omega _{en}^n \times } \right] + 2\left[ {\omega _{ie}^n \times } \right]} \right){{\bf{v}}^n},
\end{equation}
where ${\bf{T}}_b^n \in \mathbb{R}^{3 \times 3}$ is the transformation matrix from body frame  to the navigation frame. ${{\bf{f}}^b} \in \mathbb{R}^{3 \times 1}$ is the accelerometers vector state expressed in the body frame, ${{\bf{g}}^n} \in \mathbb{R}^{3 \times 1}$ is the gravity vector  expressed in the navigation frame. $\omega _{en}^n$ is the angular velocity vector between the earth centered earth fixed (ECEF) frame and the navigation frame. The angular velocity vector between ECEF and the inertial frame is given by $\omega _{ie}^n$ and the rate of change of the transformation matrix is given by \cite{farrell2008aided}
\begin{equation}
{\bf{\dot T}}_b^n = {\bf{T}}_b^n\left( {\left[ {\omega _{ib}^b \times } \right] - \left[ {\omega _{in}^b \times } \right]} \right),
\end{equation}
where $\omega _{ib}^b = {\left[ {\begin{array}{*{20}{c}}
p&q&r
\end{array}} \right]^T} \in \mathbb{R}^{3 \times 1}$ is the angular velocity vector as obtained by the gyroscope and $\omega _{in}^b$ is the angular velocity vector between the navigation frame and the inertial frame expressed in the body frame. The angular velocity between the navigation frame and the inertial frame expressed in the navigation frame is given by $\omega _{in}^n$. The alignment between body frame and navigation frame can be obtained from ${{\bf{T}}_b^n}$, as follows
\begin{equation}
{\bf \varepsilon}=\left[ {\begin{array}{*{20}{c}}
\varphi \\
\theta \\
\psi 
\end{array}} \right] = \left[ {\begin{array}{*{20}{c}}
{atan2\left( {{\bf{T}}{{_n^b}_{31}},{\bf{T}}{{_n^b}_{32}}} \right)}\\
{arccos\left( {{\bf{T}}{{_n^b}_{33}}} \right)}\\
-{atan2\left( {{\bf{T}}{{_n^b}_{13}},{\bf{T}}{{_n^b}_{23}}} \right)}
\end{array}} \right] \in \mathbb{R}^{3 \times 1},
\end{equation}
where $\varphi$ is the roll angle, $\theta$ is the pitch angle, and $\psi$ is the yaw angle. These three angles are called Euler angles.

\begin{figure}[h!]
\centering
{\includegraphics[width=0.48\textwidth]{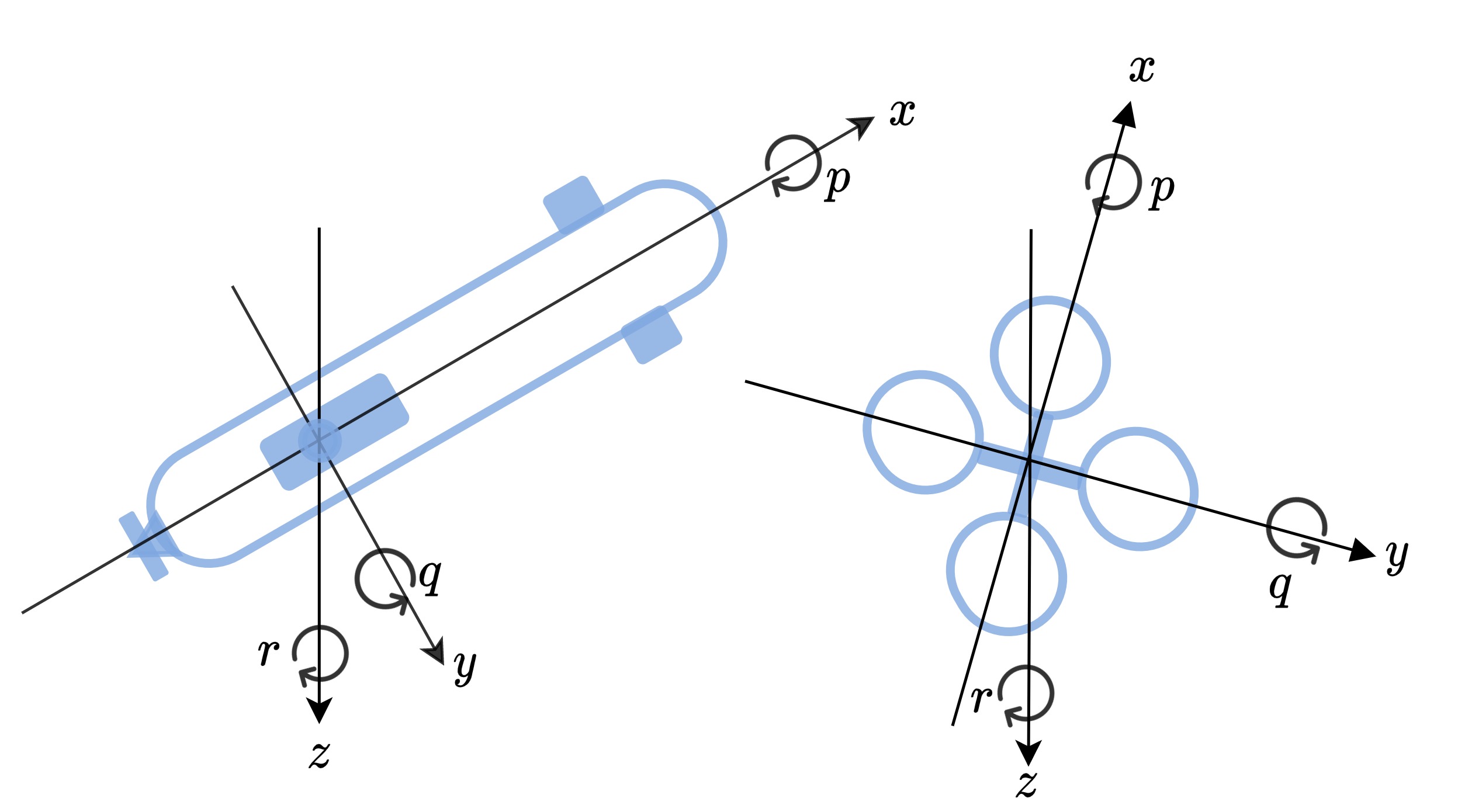}}
\caption{Autonomous underwater vehicle and quadcopter with their body axes and angular rates.}
\end{figure}

\subsubsection{IMU readings}
The GT trajectories were created by tuning vehicle angular rates, $\cal P$, and vehicle accelerations, $\bf{a}$, as follows \cite{jwo2014development}
\begin{equation}
{\cal P} = {\left[ {\begin{array}{*{20}{c}}
{\dot \varphi }&{\dot \theta }&{\dot \psi }
\end{array}} \right]^T} \in \mathbb{R}^{3 \times 1},
\end{equation}
and,
\begin{equation}
{\bf{a}} = {\left[ {\begin{array}{*{20}{c}}
{{a_x}}&{{a_y}}&{{a_z}}
\end{array}} \right]^T} \in \mathbb{R}^{3 \times 1}.
\end{equation}
The transformation between the desired motion and the INS framework for the gyroscope readings is given by \cite{jwo2014development}
\begin{equation}
{\bf{\omega }}_{ib}^b = {\bf{T}}_n^b\left( {\omega _{en}^n + \omega _{ie}^n} \right) + \overline {{\bf{\omega }}_{ib}^b} 
\end{equation}
where
\begin{equation}
\overline {{\bf{\omega }}_{ib}^b}  = \left[ {\begin{array}{*{20}{c}}
1&0&{ - \sin \theta }\\
0&{\cos \varphi }&{\sin \varphi \cos \theta }\\
0&{ - \sin \varphi }&{\cos \varphi \cos \theta }
\end{array}} \right]{\cal P} \in\mathbb{R}^{3 \times 1},
\end{equation}
and from (30) the accelerometer readings are 
\begin{equation}
{{\bf{f}}_b} = {\bf{T}}_n^b\left[ {{\bf{a}} - {{\bf{g}}^n} + \left( {\omega _{en}^n \times  + 2\omega _{ie}^n \times } \right){{\bf{v}}^n}} \right].
\end{equation}
Once, the GT IMU readings (35) and (37) are obtained, noise characteristics as described in Section III are added to create our dataset.

\bibliographystyle{IEEEtran}
\bibliography{IEEEfull}

\begin{IEEEbiography}[{\includegraphics[width=1in,height=1.25in,clip,keepaspectratio]{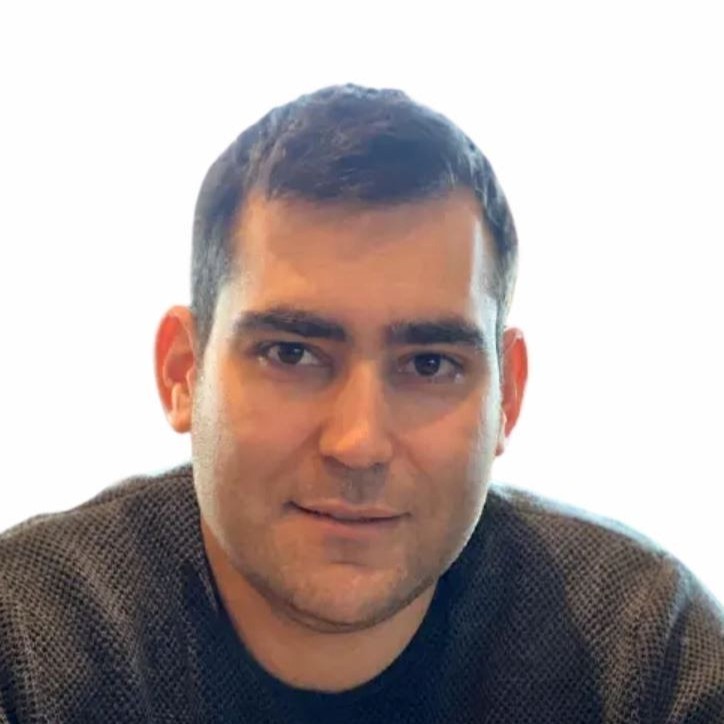}}]{Barak Or} (Member, IEEE) received a B.Sc. degree in aerospace engineering from the Technion–Israel Institute of Technology, Haifa, Israel, a B.A. degree (cum laude) in economics and management, and an M.Sc. degree in aerospace engineering from the Technion–Israel Institute of Technology in 2016 and 2018. He is currently pursuing a Ph.D. degree with the University of Haifa, Haifa. \\
His research interests include navigation, deep learning, sensor fusion, and estimation theory.
\end{IEEEbiography}

\begin{IEEEbiography}[{\includegraphics[width=1in,height=1.25in,clip,keepaspectratio]{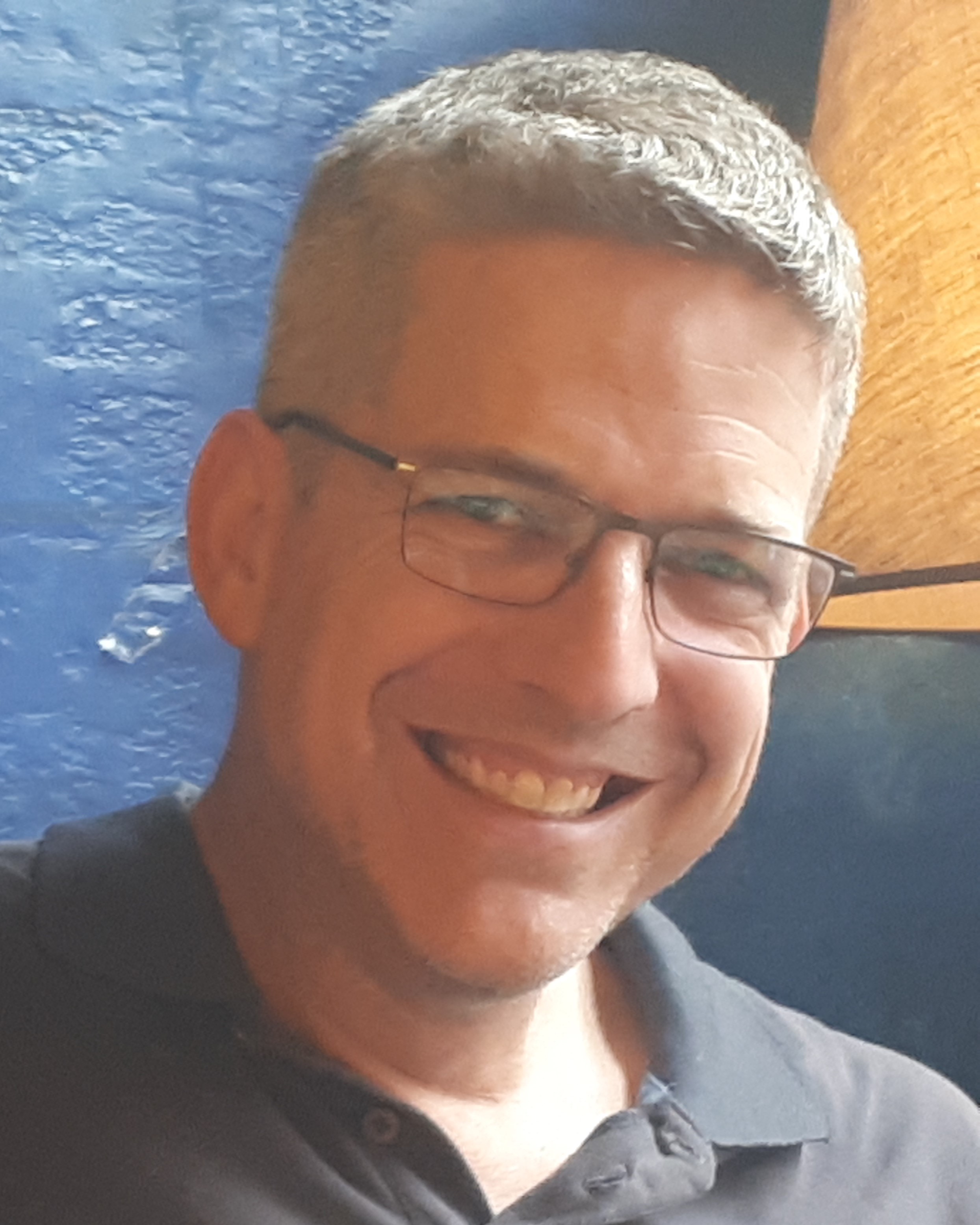}}]{Itzik Klein} (Senior Member IEEE) received B.Sc. and M.Sc. degrees in Aerospace Engineering from the Technion-Israel Institute of Technology, Haifa, Israel in 2004 and 2007, and a Ph.D. degree in Geo-information Engineering from the Technion-Israel Institute of Technology, in 2011. He is currently an Assistant Professor, heading the Autonomous Navigation and Sensor Fusion Lab, at the Hatter Department of Marine Technologies, University of Haifa. His research interests include data driven based navigation, novel inertial navigation architectures, autonomous underwater vehicles, sensor fusion, and estimation theory.
\end{IEEEbiography}

\end{document}